\documentclass{aastex631}
\usepackage{bm}
\usepackage{threeparttable}
\usepackage{amsmath}
\usepackage{graphicx}
\usepackage{hyperref}
\hypersetup{
  colorlinks=true,
  urlcolor=blue,
}

\shortauthors{Lai et al.}

\submitjournal{PSJ}

\begin{document}

\title{Ocean Circulation on Tide-locked Lava Worlds, Part I: An Idealized 2D Numerical Model}

\author[0000-0001-9700-9121]{Yanhong Lai}
\affiliation{Laboratory for Climate and Ocean-Atmosphere Studies, Department of Atmospheric and Oceanic Sciences, School of Physics, Peking University, Beijing 100871, China}
\affiliation{Institute of Ocean Research, Peking University, Beijing 100871, China}

\author[0000-0001-6031-2485]{Jun Yang}
\affiliation{Laboratory for Climate and Ocean-Atmosphere Studies, Department of Atmospheric and Oceanic Sciences, School of Physics, Peking University, Beijing 100871, China}
\affiliation{Institute of Ocean Research, Peking University, Beijing 100871, China}
\correspondingauthor{Jun Yang}
\email{junyang@pku.edu.cn}

\author[0000-0002-4615-3702]{Wanying Kang}
\affiliation{Department of Earth, Atmosphere and Planetary Science, MIT, Cambridge, MA 02139, USA}

\begin{abstract}
A magma ocean is expected to exist on the dayside of tide-locked planets if surface temperature exceeds the melting temperature of typical crust. As highly prioritized targets for the James Webb Space Telescope (JWST), more information about the surface and atmosphere of lava planets will soon be available. In most previous studies of lava planets, the system is typically assumed to be vigorously convecting and isentropic. This implies a magma ocean depth reaching $\mathcal{O}$($10^4$--$10^5$) m, determined by the adiabats and melting curves. 
In this study, we aim to simulate ocean circulation and ocean depth on tidally locked lava worlds using an idealized 2D (x-z) model developed by the authors. Our simulation results show that under zero or a small internal source, the maximum zonal current speed ranges from 0.1--1.0 m\,s$^{-1}$ and the magma ocean depth remains $\mathcal{O}$(100) m, being more than 100 times shallower than that predicted in a fully convecting system. We demonstrate that the ocean heat transport divergence is consistently smaller than the stellar insolation by 1--2 orders of magnitude. Consequently, the impact of ocean circulation on the thermal phase curve of tidally locked lava worlds is minimal in observations.
\end{abstract}


\section{Introduction}\label{sec:intro}

A hemispherical magma ocean can form and persist on tidally locked terrestrial exoplanets if the planets are very close to their host stars \citep{leger2009transiting,leger2011extreme,kite2016atmosphere,chao2021lava}. These planets, often referred to as ``lava worlds'', are more detectable due to their high substellar temperature ($T_{sub}$) and relatively large transit depth. A substantial number of lava planets have been identified\footnote{ \url{https://exoplanetarchive.ipac.caltech.edu/docs/counts\_detail.html}}, such as, Kepler-10b ($T_{sub}$ = 3000 K) \citep{batalha2011kepler,dumusque2014kepler}, CoRoT-7b ($T_{sub}$ = 2500 K) \citep{leger2009transiting}, 55 Cnc e ($T_{sub}$ = 2773 K) \citep{demory2011detection,bourrier201855}, TOI-561b ($T_{sub}$ = 3218 K) \citep{brinkman2023toi}, and K2-141b ($T_{sub}$ = 3000 K) \citep{malavolta2018ultra,nguyen2020modelling}.
Lava worlds are currently considered high-priority targets for the James Webb Space Telescope (JWST) \citep{beichman2018white}. Consequently, more observational data regarding the atmosphere and surface of lava planets will soon be available.

Owing to the extremely high surface temperature and the relatively old age of lava worlds (e.g., $\sim$2 Gyr for CoRoT-7b and $\sim$10 Gyr for Kepler-10b; \cite{leger2009transiting,batalha2011kepler}), it is possible that volatile elements (such as C, N, H) have escaped from the planets \citep{valencia2010composition}, so that the atmospheric mass will be low. The tenuous atmosphere comes from the vaporization of underlying silicate melts \citep{schaefer2009chemistry,leger2009transiting,kite2016atmosphere}. Under temperatures higher than the melting point, the magma surface initially outgases the most volatile components to the atmosphere, until atmospheric partial pressure reaches an equilibrium with the surface. 
Atmospheric pressure is determined by surface temperature, peaking at the substellar point and ranging from 10$^0$ to 10$^4$ Pa, due to the strongest evaporation \citep{leger2011extreme,nguyen2020modelling}. It gradually approaches zero at the magma ocean boundary where no evaporation occurs. Thus, there is a strong atmospheric pressure gradient, leading to strong winds flowing outward from the substellar point ($\sim$2000 m\,s$^{-1}$; \cite{castan2011atmospheres,nguyen2020modelling,kang2021escaping}).

Given the tenuous atmosphere on lava planets, the effects of atmosphere on surface temperature are generally limited and negligible \citep{ingersoll1985supersonic,leger2011extreme,kite2016atmosphere,nguyen2020modelling}. 
For molten silicates with a latent heat of vaporization $l_v \sim$ 10$^5$--10$^6$ J\,kg$^{-1}$ and a maximum evaporation rate $E \sim$ 0.1 kg\,m$^{-2}$\,s$^{-1}$ \citep{nguyen2020modelling}, a latent heat flux of 10$^4$--10$^5$ W\,m$^{-2}$ can be calculated. This is about one or two orders of magnitude smaller than the stellar insolation ($\sim$10$^6$ W\,m$^{-2}$; \cite{castan2011atmospheres}). Thus, the surface temperature of lava planets is typically assumed to be determined by local radiative balance \citep{leger2009transiting}. 

\begin{figure*}
    \centering
    \includegraphics[width=0.9\textwidth]{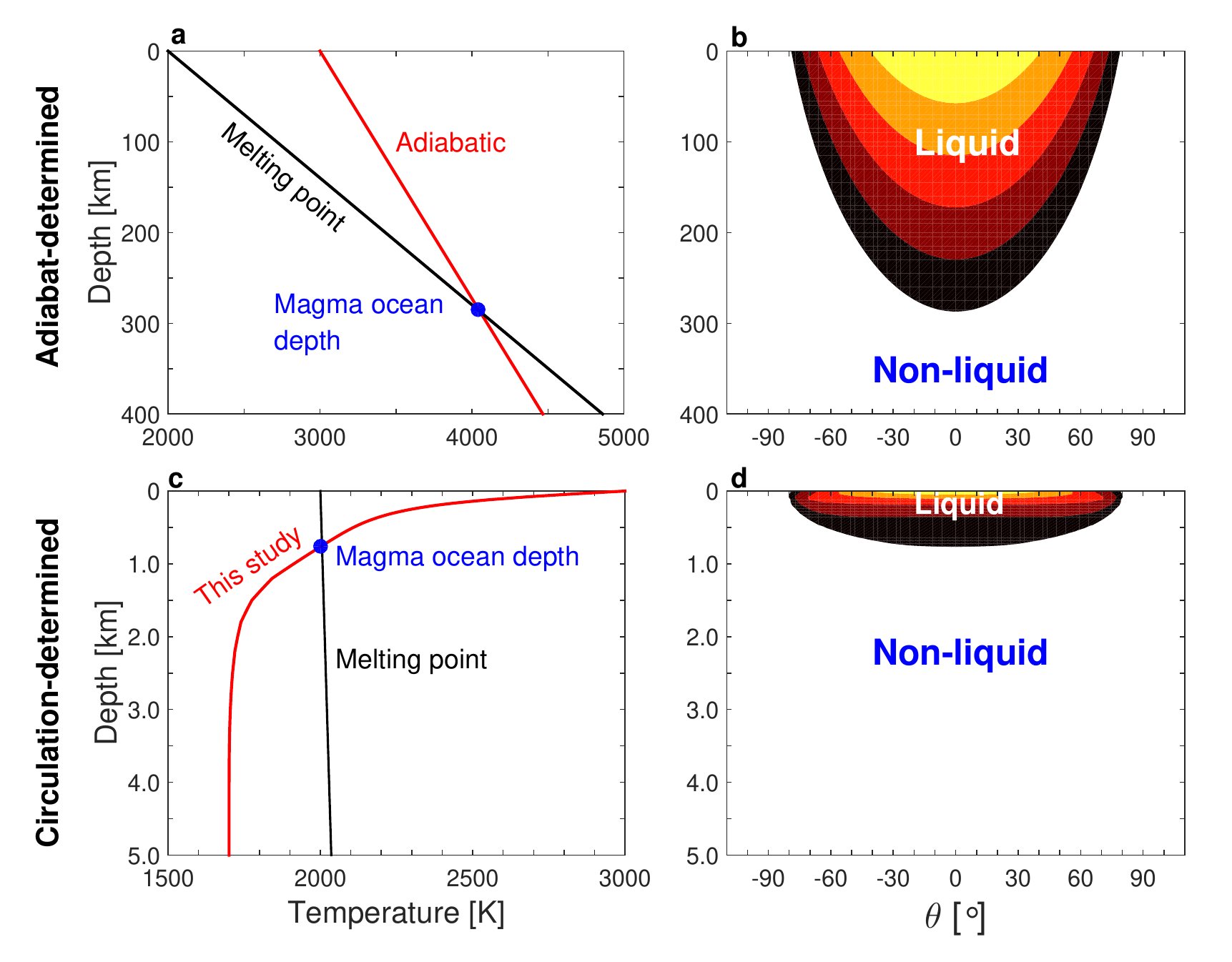}
    \caption{Vertical temperature profile and magma ocean depth when they are adiabat-determined (upper panels) and ocean circulation-determined (lower panels), respectively. (a \& c) Vertical temperature profile with a substellar temperature of 3000 K (red line) and the liquidus curve (black line). (b \& d) The magma ocean depth as a function of the angle away from the substellar point under a surface temperature shown in Figure \ref{fig:model}(b). Note that the liquidus in panel (c) increases with depth as that in panel (a), but it is not evident within the shallow depth. The vertical range is different between the upper and lower panels.}
    \label{fig:compare}
\end{figure*}

Much remains unclear regarding the ocean on lava planets. Two primary theories aim to determine the magma ocean depth on tidally locked lava planets: adiabat-determined and ocean circulation-determined. In scenarios with strong internal heating, the interior heat source may dominate over stellar radiation, leading to robust vertical convection \citep{boukare2022deep}. Vigorous convection establishes an isentropic and adiabatic vertical temperature profile, for which temperature increases with pressure following the adiabatic lapse rate (red line in Figure \ref{fig:compare}(a); \cite{solomatov2007magma,zhang2022internal}). Magma ocean depth is determined by the intersection of this adiabatic temperature profile and the melting curve (black line in Figure \ref{fig:compare}(a); \cite{herzberg1996melting,fiquet2010melting,leger2011extreme}). 
Consequently, the magma ocean depth can reach tens to hundreds of kilometers (Figure \ref{fig:compare}(a \& b); \cite{leger2011extreme,boukare2022deep,meier2023,boukare2023}).

The interior temperature profile can become sub-adiabatic (i.e., the vertical temperature gradient is smaller than the adiabatic lapse rate), when the internal heat source is weak or absent. Under such conditions, the magma ocean may not exhibit vigorous convection. \cite{meier2023} utilized 2D mantle convection models to investigate the potential magma ocean depth of 55 Cnc e under varying internal heating rates. Their findings indicate that the magma ocean depth determined by a sub-adiabatic temperature profile is significantly shallower compared to that governed by adiabatic and super-adiabatic profiles. Nevertheless, the magma ocean depth still reaches approximately 500 km on the day side, even without internal heating. The effect of ocean circulation on magma ocean depth was not included in their study.

Magma ocean depth will be governed by ocean circulation when the internal heat source is weak or absent. This scenario is possible for tidally locked lava planets, especially those that have cooled over gigayears (such as CoRoT-7b, Kepler-10b, 55 Cnc e; \cite{leger2011extreme,batalha2011kepler,von201155,malavolta2018ultra,brinkman2023toi}).
The overturning circulation is possible on lava planets, given that the viscosity of molten silicates is comparable to that of seawater in Earth’s oceans \citep{dingwell2004viscosity,haynes2014crc,sun2020physical,zhang2022internal}. With a large surface temperature gradient and strong atmospheric winds \citep{leger2011extreme,castan2011atmospheres}, thermal-driven and wind-driven overturning circulation similar to Earth's oceans could exist within the magma ocean. As suggested by \cite{kite2016atmosphere}, magma ocean depth in the presence of a thermal-driven overturning circulation might be over 100 times shallower than that determined by the adiabats. However, their magma ocean depth was obtained using scaling analysis, and numerical simulations for how ocean circulation reduces magma ocean depth are lacking. Furthermore, the effect of wind forcing was not taken into account. Numerical simulations of magma oceans on tidally locked lava planets are strongly needed.

To simulate ocean circulation on lava planets, general circulation models (GCMs), such as MITgcm \citep{Marshall1997a,Marshall997b}, face three main challenges. Firstly, the ocean depth, marking the boundary between liquid and non-liquid regions, continuously evolves with temperature until equilibrium is reached. Secondly, owing to the extensive temperature range on lava planets, silicates exist in three different states: solid, partially molten, and liquid. Thus, it is necessary to account for the variation of silicate viscosity and diffusivity with temperature \citep{dingwell2004viscosity,ghosh2011diffusion,ni2015transport}. In current GCMs, only constant values or prescribed spatial patterns for viscosity and diffusivity are available. Thirdly, silicate viscosity also changes drastically with melt fraction, varying from $10^{18}$ to $10^{-4}$ m$^2$\,s$^{-1}$ from the solidus (melt fraction is zero) to the liquidus (melt fraction is 100\%; see Section \ref{vis_dif}). The large value and variation of viscosity can easily lead to numerical instability. 

Here we simulate the ocean overturning circulation on tidally locked lava worlds using an idealized two-dimensional (2D, x-z) model that we developed. Note that only a small internal heat source (0.5 W\,m$^{-2}$, corresponding to a temperature of 50 K on the night side; \cite{leger2011extreme}) is included in this study, given that most tidally locked lava worlds have cooled over billions of years.
We present detailed model descriptions and experimental designs in Section \ref{sec:model}. In Section \ref{thermal}, the ocean circulation and magma ocean depth with thermal forcing only are shown. Section \ref{wind} presents simulation results with both thermal and wind forcings. The comparison between our study and previous studies is discussed in Section \ref{compare}. Implications of our results for future observations are discussed in Section \ref{observe}. In Section \ref{sec:discuss}, we discuss the potential influence of varying factors on the simulation results. In Section \ref{sec:conclude}, we summary the results.

It is important to note that the magma ocean investigated in this study differs from the global magma ocean formed during the formation of rocky planets, such as early Mars and early Earth \citep{elkins2012magma,monteux2016cooling}. The magma ocean on tidally locked lava planets is long-lived owing to persistent stellar insolation. In contrast, the cooling and crystallization of the magma ocean formed in the early planetary evolution is a rapid process, with a solidification time ranging from thousands to millions of years \citep{hamano2015lifetime,monteux2016cooling,nikolaou2019factors}.

\section{Model Descriptions and Experimental Designs}\label{sec:model}
We simulate the magma ocean using a self-developed 2D (x-z) model programmed in Matlab 2019a. For simplicity, the y direction is not resolved, and the meridional velocity is assumed to be zero everywhere. The effect of planetary rotation is also excluded in the 2D model. Referring to the 3D equations employed in MITgcm (a global ocean model for oceans on Earth; \cite{Marshall1997a,Marshall997b}), the 2D governing equations for the magma ocean on lava worlds in the Cartesian coordinate can be written as follows:
\begin{equation}
    \frac{\partial u} {\partial t} + u \frac{\partial u} {\partial x} + w \frac{\partial u} {\partial z}  = - \frac{1} {\rho_c} \frac{\partial p} {\partial x} + \frac{\partial} {\partial x} (A_h \frac{\partial u} {\partial x}) + \frac{\partial} {\partial z} (A_z \frac{\partial u} {\partial z})   + F_u,
\label{equat1}
\end{equation}
\begin{equation}
    \frac{\partial \theta} {\partial t} + u \frac{\partial \theta} {\partial x} + w \frac{\partial \theta} {\partial z}  =  \frac{\partial} {\partial x} (k_h \frac{\partial \theta} {\partial x}) +  \frac{\partial} {\partial z} (k_z \frac{\partial \theta} {\partial z}) + F_{\theta},
\label{equat2}
\end{equation}
\begin{equation}
\frac{\partial u} {\partial x} + \frac{\partial w} {\partial z} = 0,
\label{equat3}
\end{equation}
\begin{equation}
\frac{\partial \eta} {\partial t} + \frac{\partial H \hat{u}} {\partial x} = 0,
\label{equat4}
\end{equation}
\begin{equation}
T = T (\theta, p),
\label{equat5}
\end{equation}
\begin{equation}
\rho = \rho (T, p); \rho_s = \rho (T_s, 0),
\label{equat6}
\end{equation}
\begin{equation}
p = \rho_s g \eta +\int_z^0{\rho g}dz,
\label{equat7}
\end{equation}
\begin{equation}
A_h=A_h(T); A_z=A_z(T),
\label{equat8}
\end{equation}
where $x$ is the horizontal distance from the substellar point; $z$ is the vertical distance from surface and is negative; $u$ and $w$ are zonal and vertical current speeds, respectively; $\theta$ and $T$ are potential temperature and temperature, respectively; $\rho$ is ocean density; $\rho_c$ is the reference density and a value of 2600 kg\,m$^{-3}$ is used; $T_s$ and $\rho_s$ are temperature and density at the surface, respectively; $\eta$ is sea surface height (SSH), which is determined by the vertical integral of the horizontal divergence of the zonal velocity ($\frac{\partial H\hat{u}}{\partial x}$, where $H\hat{u} = \int_{-H}^{\eta} u dz$ and $H$ is the depth of the vertical domain); $g$ is gravity; $p$ is pressure, which is composed of a barotropic part due to variations in surface height ($\eta$) and a hydrostatic part due to the vertical integral of density ($\rho$); $A_h$ and $A_z$ are horizontal and vertical viscosity coefficients, respectively; $k_h$ and $k_z$ are horizontal and vertical diffusivities, respectively; and $F_u$ and $F_{\theta}$ are zonal wind forcing and surface temperature forcing, respectively. 

In a multicomponent system \citep{schaefer2009chemistry,zilinskas2022observability}, silicates exhibit different states based on temperature \citep{stevenson2010planetary}. They are solid when the temperature is below the solidus, partially molten between the solidus and liquidus, and fully molten when the temperature is above the liquidus. 
Partially molten silicates exhibit solid-like behavior when the crystal fraction exceeds a critical threshold, while behaving more fluid-like when the crystal fraction falls below this threshold. This critical value is highly dependent on the rheology (e.g., the crystal composition, size, and shape, etc) \citep{lejeune1995rheology,rintoul1996computer,costa2005viscosity,zhang2022internal}.
Both the solidus and liquidus increase with increasing pressure \citep{monteux2016cooling,zhang2022internal}. Due to the shallow depths (see below), their variations are limited. Therefore, both the solidus ($T_{sol}$) and liquidus ($T_{liq}$) are considered as constants in this study. By default, $T_{sol}$ = 1700 K and $T_{liq}$ = 2000 K \citep{monteux2016cooling}. 
It is worth mentioning that the melting temperature of silicates rises if the solid interior of the planet is depleted in volatiles, which might threaten the persistence of the magma ocean \citep{gelman2011}. This scenario could occur on tidally locked lava worlds, given that their magma oceans have vaporized over billions of years. 

\begin{figure*}
    \centering
    \includegraphics[width=0.95\textwidth]{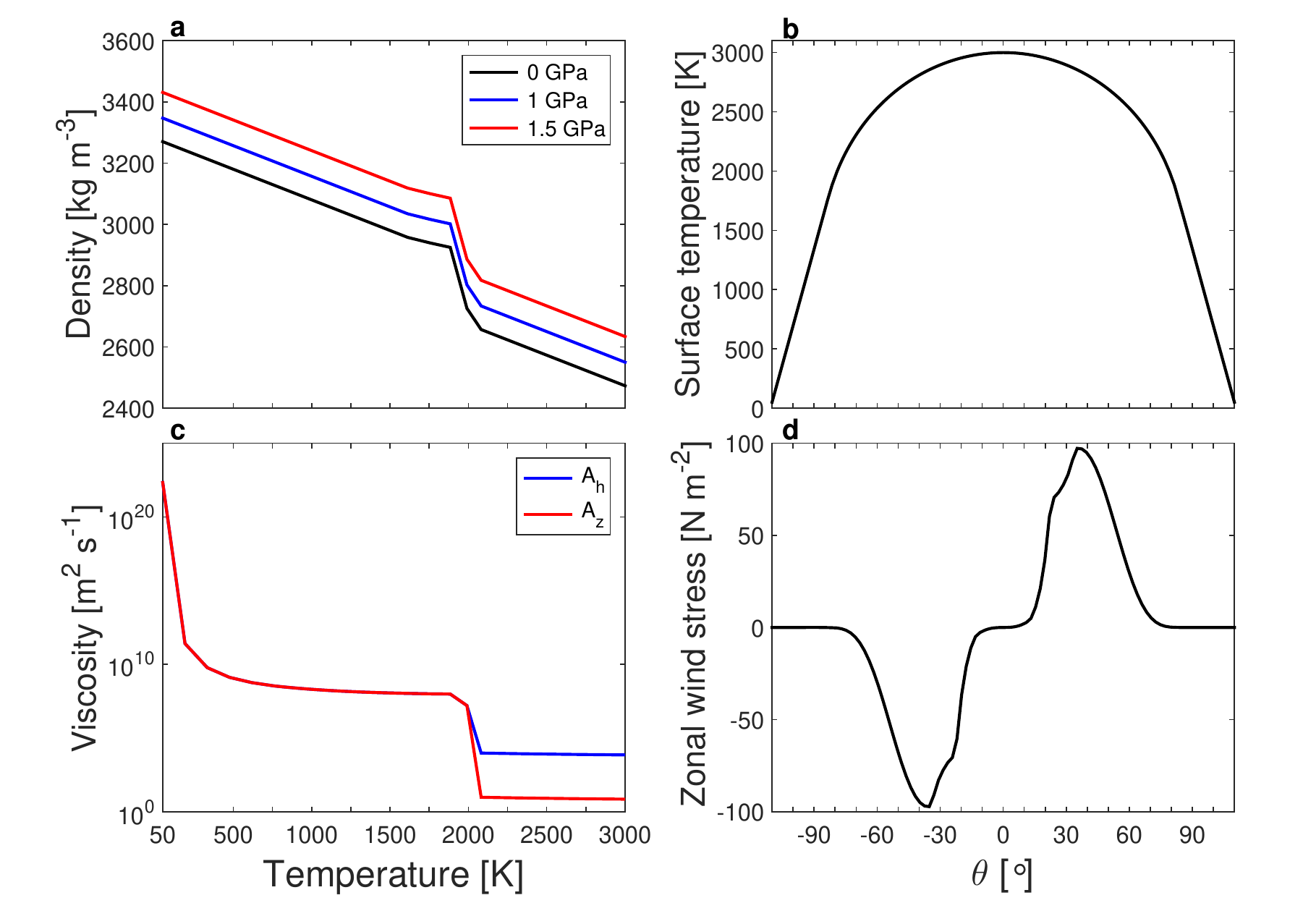}
    \caption{(a \& c) Equation of state used in the model (a); horizontal ($A_h$) and vertical ($A_z$) viscosity coefficients as a function of temperature (c). (b \& d) External forcings imposed at the surface. Surface temperature with a substellar temperature of 3000 K and a temperature of 50 K at $\pm$110$^{\circ}$ (b); zonal wind stress with a maximum value of $\sim$100 N\,m$^{-2}$ near 40$^{\circ}$, positive (negative) value corresponds to eastward (westward) wind stress (d).}
    \label{fig:model}
\end{figure*}

It is important to note that Equation (\ref{equat2}) predicts potential temperature rather than the absolute temperature. To determine silicate density and viscosity (Equations (\ref{equat6}) and (\ref{equat8})), a conversion from potential temperature to temperature following the adiabatic lapse rate is required (Equation (\ref{equat5}); \cite{fofonoff1962physical,mcdougall2003causes}). The difference between potential temperature and temperature is limited, compared to the temperature itself. 

\subsection{Equation of state}\label{eos}
The equation of state characterizes the dependence of silicate density on temperature and pressure, typically determined using the third-order Birch-Murnaghan equation of state (EoS) \citep{sakamaki2010density,katsura2010adiabatic}. 
To simplify, linear and quadratic approximations are employed to represent the relationship between density and temperature and between density and pressure, respectively (Equation (\ref{equat6}) and Figure \ref{fig:model}(a)). These approximations are based on fitting the third-order Birch-Murnaghan EoS.
To calculate density, the densities at the liquidus ($\rho_{l0}$) and at the solidus ($\rho_{s0}$) at zero pressure are defined, respectively,
\begin{equation}
\rho_{l0} = \rho_0 + \alpha_1 (T_{liq} - T_0), \rho_{s0} = \rho_{l0} (1+\Delta \rho),
\label{equat9}
\end{equation}
where $\rho_0$ = 2.673 g\,cm$^{-3}$ is the reference density at a reference temperature ($T_0$ = 2000 K) and at zero pressure, $\alpha_1 = -2\times 10^{-4}$ g\,cm$^{-3}$\,K$^{-1}$ is the slope of density with temperature, $\Delta \rho$ = 10\% is the density difference between the solidus and liquidus \citep{tosi2013overturn,ghosh2016solid}. Then, the interior density can be given as, 
\begin{equation}
\rho = \left\{
\begin{array}{rcl}
\rho_s = \rho_{s0}+\alpha_1(T-T_{sol}) + \beta_1 p^2 + \beta_2 p, & &  {T < T_{sol}}\\
\rho_t = (\rho_{s0} + \beta_1 p^2 + \beta_2 p ) \cdot \tanh(c_1-(c_1 - c_2) \frac{T-T_{sol}} {T_{liq} - T_{sol}} ), & & {T_{sol} \leq T < T_{liq}}\\
\rho_l = \rho_{l0}+\alpha_1(T-T_{liq}) + \beta_1 p^2 + \beta_2 p, & & {T \geq T_{liq}}
\end{array}\right.
\label{equat10}
\end{equation}
where $\rho_s$, $\rho_t$, and $\rho_l$ are the densities in solid, transitional, and liquid states, respectively; $\beta_1$ and $\beta_2$ are constant parameters determining the variation of density with pressure, with values of 0.00667 g\,cm$^{-3}$\,GPa$^{-2}$ and 0.1022 g\,cm$^{-3}$\,GPa$^{-1}$, respectively; $c_1$ and $c_2$ are parameters set to ensure the continuity of density with temperature in different temperature ranges, satisfying $\tanh(c_1)$ = 1 and ${\rho_l (T = T_{liq})} /{\rho_s (T = T_{sol})}$ = atanh($c_2$), where $c_2$ is non-constant and varies with pressure. Note that $\rho$ is in g\,cm$^{-3}$ and pressure is in GPa (10$^9$ Pa) in Equations (\ref{equat9}) and (\ref{equat10}). Equation (\ref{equat10}) and Figure \ref{fig:model}(a) show that density decreases linearly with increasing temperature when the temperature is below the solidus or above the liquidus. In the range between the solidus and liquidus, density decreases with temperature following a hyperbolic tangent (tanh) function. Furthermore, density always increases quadratically with increasing pressure.

\subsection{Viscosity and diffusivity}\label{vis_dif}
Silicate viscosity is dependent on pressure and temperature approximately following an Arrhenius relationship \citep{dingwell2004viscosity,liebske2005viscosity,ghosh2011diffusion}. The effect of pressure on silicate viscosity is relatively small and can be ignored \citep{liebske2005viscosity}. Silicate viscosity also changes drastically with melt fraction, ranging from 10$^{18}$ to 10$^{-4}$ m$^2$\,s$^{-1}$ from the solidus to liquidus \citep{sun2020physical,zhang2022internal}. Given the high viscosity at the solidus, it is likely that the current speed in solid regions is extremely small and can be approximated as zero. The significant variation of viscosity below the liquidus plays a critical role in ocean dynamics. As mentioned in Section \ref{sec:intro}, this is also the reason why we choose to develop an idealized 2D model instead of employing a 3D ocean general circulation model. 

To simplify, silicate viscosity can be regarded as a function of temperature following an exponential decay. Between the solidus and liquidus, it increases rapidly with decreasing temperature following a hyperbolic tangent (tanh) function (Figure \ref{fig:model}(c)). For example, vertical viscosity is expressed as,
\begin{equation}
A_z = \left\{
\begin{array}{rcl}
A_{zsol} \cdot e^{-\frac{T-T_{sol}} {T}}, & &  {T < T_{sol}}\\
A_{zsol} \cdot \tanh(a-(a - b) \frac{T-T_{sol}} {T_{liq} - T_{sol}} ), & & {T_{sol} \leq T < T_{liq}}\\
A_{zliq} \cdot e^{-\frac{T-T_{liq}} {T}} , & & {T \geq T_{liq}}
\end{array}\right.
\label{equat11}
\end{equation}
where $A_{zsol}$ and $A_{zliq}$ are the vertical viscosity at the solidus and liquidus, respectively; both $a$ and $b$ are constants 
ensuring a continuous change in viscosity with temperature at different temperature ranges. By default, $A_{zsol}$ = 10$^8$ and $A_{zliq}$ = 10$^1$ m$^2$\,s$^{-1}$. 
Similarly, the variation of horizontal viscosity with temperature follows Equation (\ref{equat11}) with $A_{hsol}$ = 10$^8$ and $A_{hliq}$ = 10$^4$ m$^2$\,s$^{-1}$.
Between the solidus and liquidus, the strict relationship between the viscosity of partially molten silicates and the melt fraction is not well considered \citep{monteux2016cooling,zhang2022internal}. Under the tanh function (Equation (\ref{equat11})), zonal velocity of partially molten silicates decreases significantly when the temperature falls below 1900 K (Section \ref{sec:result}). For simplicity, the liquidus is directly used to determine the magma ocean boundary. Note that solid network may be compacted at the bottom of the magma ocean under high gravity, leading to the result that the silicate viscosity can be increased by tens of orders of magnitude when the temperature just falls below the liquidus \citep{solomatov2007magma}. Under this circumstance, the velocity will decrease significantly when the temperature just falls below 2000 K, which may result in a relatively shallower magma ocean.  

The horizontal and vertical viscosities at the solidus are lower than the realistic values ($\sim$10$^{18}$ m$^2$\,s$^{-1}$; \cite{zhang2022internal}), due to computational efficiency. Nonetheless, this modification does not influence the simulation results within the ocean, as long as the viscosity below the solidus sufficiently reduces velocity to nearly zero.
The horizontal and vertical viscosities at the liquidus are considerably higher than the molecular viscosity of silicate melts ($\sim$10$^{-4}$ m$^2$\,s$^{-1}$; \cite{zhang2022internal}). In addition, the horizontal viscosity exceeds the vertical viscosity by three orders of magnitude. 
These values of viscosity at the liquidus are chosen for two primary reasons. Firstly, the viscosity of silicate melts should be dominated by eddy viscosity rather than molecular viscosity in the presence of circulations and turbulences \citep{li2017lateral,luo2023study,sentchev2023estimation}. 
Generally, the horizontal eddy viscosity is higher than the vertical eddy viscosity, given that the horizontal scale is much larger than the vertical scale \citep{prandtl19257,heisel2020}.
Secondly, numerical simulations require a minimum viscosity threshold for stability. Due to the absence of the Coriolis force in our 2D model, current speed will increase rapidly with decreasing viscosity. At the molecular viscosity level, the current speed may exceed tens of meters per second, destabilizing the 2D model. 

Similar to silicate viscosity, the molecular diffusivity of molten silicates also depends on pressure and temperature roughly following the Arrhenius relation \citep{gibert2003thermal,ghosh2011diffusion,ni2015transport}. For simplicity, the influence of pressure on molecular diffusivity is relatively limited and can be neglected. The molecular diffusivity of molten silicates can vary from 10$^{-9}$ to 10$^{-8}$ m$^2$\,s$^{-1}$ as the temperature increases from 2000 to 3000 K \citep{ghosh2011diffusion}.
In contrast, the molecular diffusivity of solid silicates is significantly smaller, sometimes reaching values as low as 10$^{-20}$ m$^2$\,s$^{-1}$ for certain chemical compositions \citep{zhang2010diffusion,ni2015transport}. However, eddy diffusions within the ocean, generated by wave breaking, winds, and tides, are more important. In Earth’s oceans, horizontal and vertical eddy diffusivities typically range from 10$^2$ to 10$^4$  m$^2$\,s$^{-1}$ \citep{cole2015eddy} and from 10$^{-5}$ to 10$^{-3}$ m$^2$\,s$^{-1}$ \citep{munk1998abyssal,Waterhouse2014}, respectively. These values are significantly higher than molecular diffusivity of seawater (10$^{-7}$--10$^{-6}$ m$^2$\,s$^{-1}$; \cite{wunsch2004}). Similar to Earth, we assume that eddy diffusivity dominates diffusion within the ocean, and constant values are used. By default, $k_h$ = 10$^3$ m$^2$\,s$^{-1}$, and $k_z$ = 10$^{-4}$ m$^2$\,s$^{-1}$. 

\begin{table}
\caption{Planetary and Oceanic Parameters in the Control Experiment \label{tab1}}
\centering
\begin{tabular}{lll}
\hline
Parameter  & Value  & Units\\
\hline
Planet radius ($a$) & 9000 & km\\
Vertical depth ($H$) & 5000 & m\\
Planet gravity ($g$) & 22 & m\,s$^{-2}$\\
Thermal expansion coefficient ($\alpha$) & 2$\times 10^{-4}$ & K$^{-1}$\\
Heat capacity at constant pressure ($c_p$) & 1800 & J\,kg$^{-1}$\,K$^{-1}$\\
Density contrast ($\Delta \rho$) & 10$\%$\\
Substellar temperature ($T_{sub}$)  & 3000 & K \\
Solidus ($T_{sol}$) & 1700 & K\\
Liquidus ($T_{liq}$) & 2000 & K\\
Horizontal and vertical viscosities at solidus & 10$^8$, 10$^8$ & m$^2$\,s$^{-1}$\\
Horizontal and vertical viscosities at liquidus & 10$^4$, 10$^1$ & m$^2$\,s$^{-1}$ \\
Horizontal diffusivity ($k_h$) & $10^3$ & m$^2$\,s$^{-1}$\\
Vertical diffusivity ($k_z$) & $10^{-4}$ & m$^2$\,s$^{-1}$\\
\hline
\multicolumn{3}{l}{}
\end{tabular}
\end{table}

Beyond the ocean, diffusivity should be dominated by molecular diffusivity, which is significantly lower than that within the ocean. 
In our simulations, however, both horizontal and vertical diffusivities are globally uniform for two reasons. Firstly, variable diffusivity over time and space can easily lead to numerical instability due to the limitations of the 2D model. Secondly, the time required for the system to reach equilibrium is inversely related to the magnitude of diffusivity. Using an extremely low diffusivity would greatly extend the time needed to attain equilibrium (Appendix \ref{ap:timescale_momentum}). Under this approximation, horizontal diffusivity is greater than vertical diffusivity by several orders of magnitude even outside the ocean.  This significant difference is partially responsible for the non-constant vertical temperature profile observed outside the ocean (see Sections \ref{thermal} \& \ref{wind}).

\subsection{External forcings}\label{forcing}
The ocean model has not been coupled to the atmosphere, but forced by surface temperature and wind stress, $F_{\theta}$ and $F_u$. Surface temperature (equal to potential temperature at the surface) is restored to a prescribed distribution, $\theta^{\ast}$ (Figure \ref{fig:model}(b)), with a relaxation timescale ($\tau_{\theta}$) of 10 Earth days,
\begin{equation}
F_{\theta} = - \frac{1}{\tau_{\theta}} (\theta - \theta^{\ast}).
\label{equat12}
\end{equation}
For example, in the eastern hemisphere: $\theta^{\ast}$ follows the local radiative balance from 0$^{\circ}$ to 85$^{\circ}$ \citep{leger2011extreme,castan2011atmospheres}, and the substellar temperature of Kepler-10b is adopted (T$_{sub}$ = 3000 K; \cite{batalha2011kepler}); between 85$^{\circ}$ and 110$^{\circ}$, $\theta^{\ast}$ decreases linearly with the angle away from the substellar point to 50 K (induced by the possible geothermal heating of 0.5 W\, m$^{-2}$ on the nightside; \cite{leger2011extreme}). Regions beyond 110$^{\circ}$ are not resolved, in which magma ocean will not form.

Zonal wind forcing is given as
\begin{equation}
F_u = \frac{\tau_x} {\rho_c \Delta z_s},
\label{equat13}
\end{equation}
where $\tau_x$ is the zonal wind stress and $\Delta z_s$ is the depth of the surface layer. Following \cite{ingersoll1985supersonic}, zonal wind stress on lava worlds can be given as $\tau_x = \rho_a \omega_a V$, where $\rho_a$ is the atmospheric density in the boundary layer, $V$ is the atmospheric wind speed, $\omega_a$ is the momentum transfer coefficient between surface and atmosphere. $\omega_a$ is given by
\begin{equation}
\omega_a = \left\{
\begin{array}{rcl}
\frac{V_e^2 - 2 V_d V_e + 2 V_d^2} {-V_e + 2V_d}, & &  {V_e < 0}\\
\frac{2 V_d^2} {V_e + 2V_d} , & & {V_e \ge 0}
\end{array}\right.
\label{equat14}
\end{equation}
where $V_e = \frac{mE}{\rho_a}$, which represents the contribution of the mean flow advection normal to the surface on the momentum exchange, $E$ is the evaporation rate, $m$ is the mass per molecule; $V_d = \frac{V_{\ast}^2} {V} + C_d V$, which represents the contribution of the turbulent eddies on the momentum transfer, $V_{\ast}$ is the friction velocity and $C_d$ is the surface drag coefficient. 
The atmospheric wind speed, density, temperature, and evaporation rate of Kepler-10b with a SiO atmosphere are adopted \citep{kang2021escaping}. With strong wind speed on lava planets ($\sim$2000 m\,s$^{-1}$) , $C_d$ is taken as 10$^{-2}$, which is slightly higher than the estimate on Earth ($\sim$2$\times$10$^{-3}$ under wind speed of 40 m\,s$^{-1}$; \cite{sterl2017drag,jiang2021high}). Figure \ref{fig:model}(d) shows that the zonal wind stress is zero at the substellar point, and reaches its peak, around $\pm$100 N\,m$^{-2}$, at approximately $\pm$40$^{\circ}$. Note that the possible amplitude of zonal wind stress on lava planets is $\sim$1000 times greater than that on Earth ($\sim$0.1 N\,m$^{-2}$; \cite{risien2008global}).

\subsection{Experimental Designs and Numerical Schemes}\label{design}

In the control simulation, only thermal forcing is included (Figure \ref{fig:model}(b)). The planetary parameters (including planet radius and gravity) of Kepler-10b are adopted \citep{dumusque2014kepler}. The planetary and oceanic parameters for the control experiment are summarized in Table \ref{tab1}. Cases with both thermal and wind forcings included are also examined (Figure \ref{fig:model}(b \& d)). Note that, due to the influence of strong wind stress, the simulation becomes unstable when the default value of vertical viscosity (10$^1$ m$^2$\,s$^{-1}$) is employed. Thus, for the simulations with strong wind forcing included, vertical viscosity at the liquidus is set to 10$^2$ m$^2$\,s$^{-1}$. By default, a 5-km-deep and flat-bottomed depth is adopted. 

\begin{figure*}
    \centering
    \includegraphics[width=0.99\textwidth]{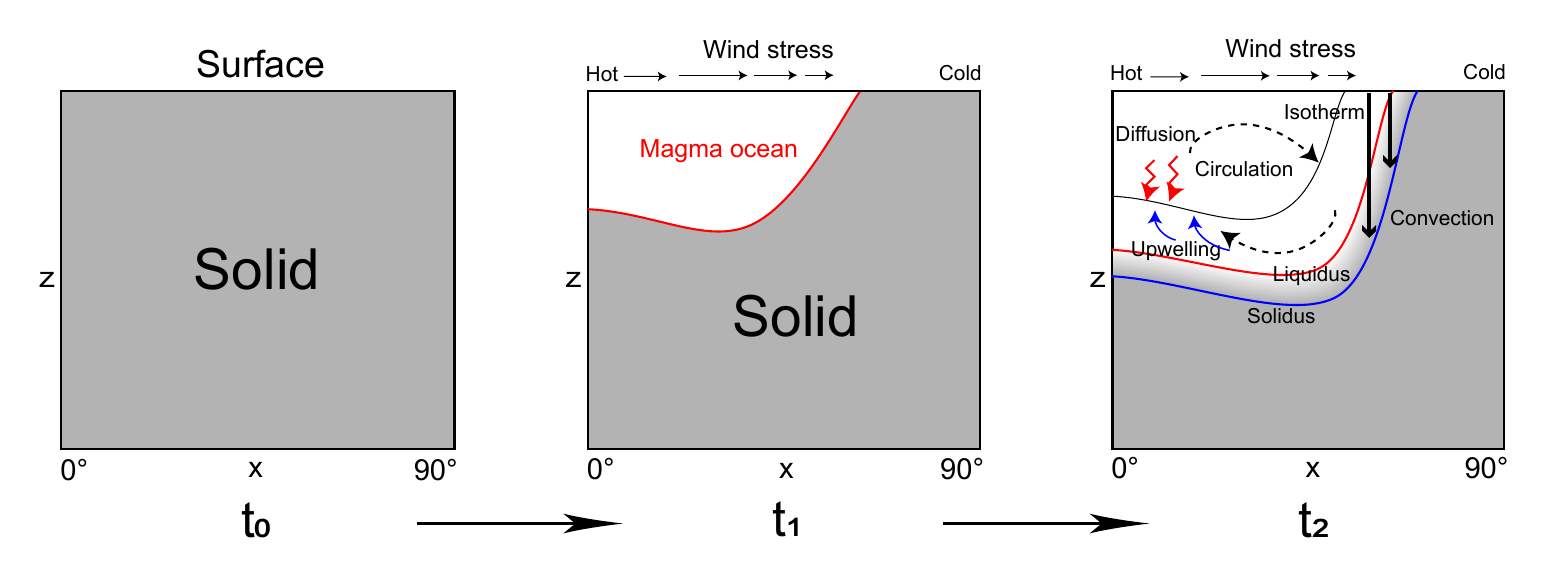}
    \caption{Schematics of the evolution of magma ocean under thermal and wind forcings. t$_0$, t$_1$, t$_2$ correspond to the initial state, the intermediate state, and the final equilibrium state, respectively. The determination processes of magma ocean depth are shown in the equilibrium state.}
    \label{fig:schematic}
\end{figure*}

The horizontal resolution is set to 2.2$^{\circ}$ (360 km). 
Vertically, we use 49 unequally spaced levels, and the vertical resolution varies nonlinearly from 40 m at the surface to 400 m at the bottom layer. 
At the surface layer, vertical velocity is determined by the evolution of sea surface height over time ($\frac{\partial \eta}{\partial t}$), which becomes zero when the system reaches equilibrium. At the bottom layer, vertical velocity is set to zero, and zonal velocity is strongly damped to zero using a linear drag with a coefficient of 10$^3$ s$^{-1}$.
No-flux boundary condition is used for temperature there.
The differential equations are solved using the finite difference method. In space discretization, a centered difference method with second-order accuracy is used. For simplicity, all variables are written on the same grids, and staggered grids are not employed. In time differentiation, a backward implicit scheme is used to solve the viscosity terms of Equation (\ref{equat1}) for numerical stability. For the remaining terms of Equation (\ref{equat1}) and all terms in Equations (\ref{equat2}) and (\ref{equat4}), a forward explicit scheme is employed. 

All simulations are initialized from a state of rest and a solid phase (Figure \ref{fig:schematic}).
The initial temperature is set to closely resemble the temperature structure dominated by vertical diffusion. Specifically, in areas where the surface temperature falls below the solidus (1700 K), the initial interior temperature is set to be equal to the surface temperature. 
Meanwhile, in regions where the surface temperature surpasses the solidus, both the surface and interior temperatures are set to the solidus. 
Notably, the equilibrated results of simulations are independent of the initial states.

\section{Results}\label{sec:result}

\subsection{Ocean Circulation under Thermal Forcing}\label{thermal}

\begin{figure*}
    \centering
    \includegraphics[width=0.95\textwidth]{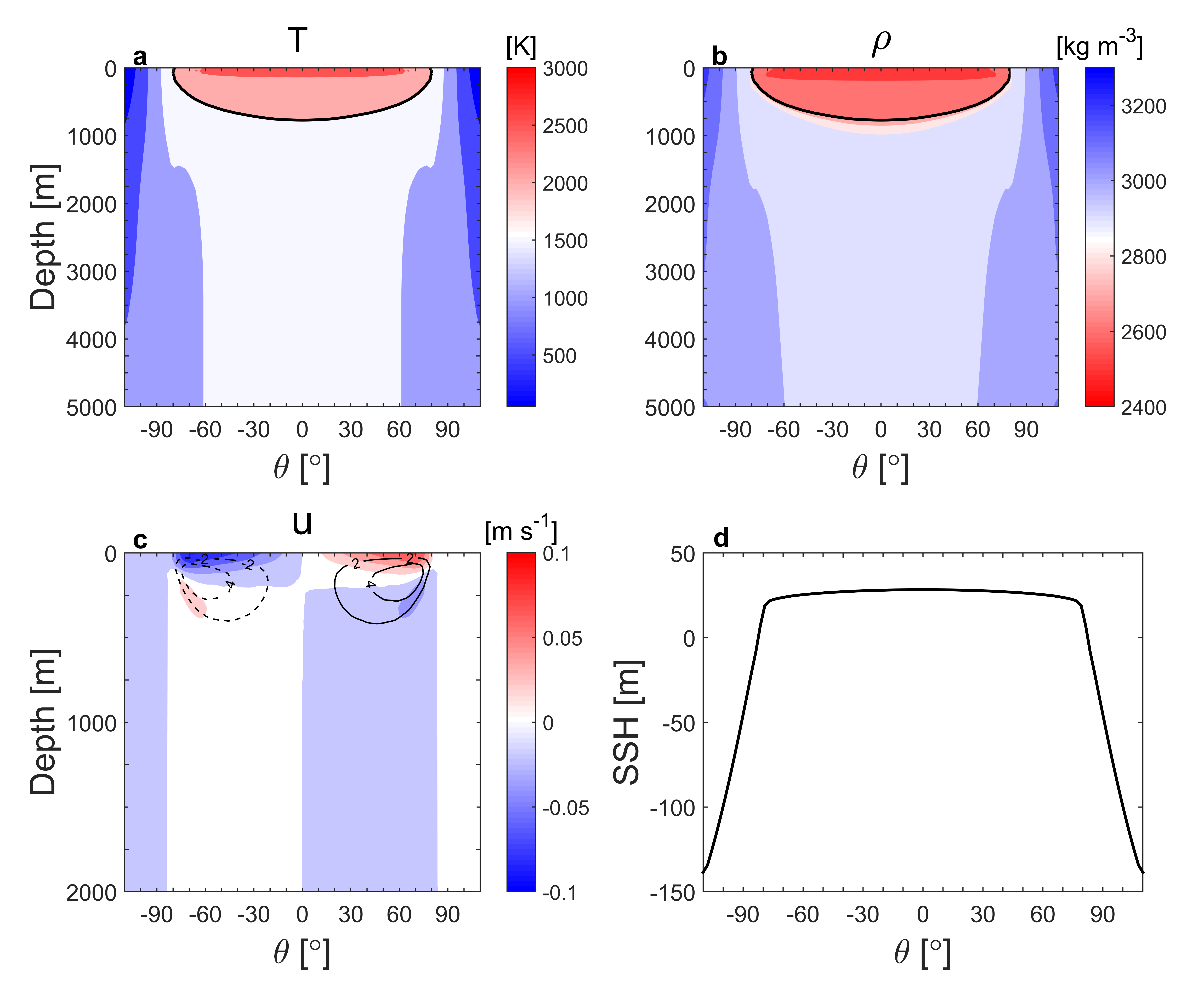}
    \caption{Snapshot results of the simulation with thermal forcing only. (a-d) Temperature ($T$) in K, density ($\rho$) in kg\,m$^{-3}$, zonal current speed ($u$) in m\,s$^{-1}$ (colors) and velocity stream function ($\psi$) in m$^2$\,s$^{-1}$ (contours) with an interval of 2.0 m$^2$\,s$^{-1}$, and sea surface height (SSH) in m. The black solid lines in panels (a)-(b) indicate the location of magma ocean boundary, which is determined by the liquidus. Solid lines in panel (c) correspond to positive values (a clockwise overturning circulation), and dashed lines correspond to negative values (an anti-clockwise overturning circulation). As the ocean circulation is confined to the upper $\sim$1 km, panel (c) shows only the zonal velocity within the upper 2 km.}
    \label{fig:thermal}
\end{figure*}

Magma ocean depth in the presence of thermal-driven overturning circulation is only several hundreds of meters deep (Figure \ref{fig:thermal}). Initialized with a completely solid state and driven by thermal forcing, the system reaches statistical equilibrium after $\sim$600 years of integration (Figure \ref{fig:time-series}(a-c)). Surface temperature, determined by the thermal forcing shown in Figure \ref{fig:model}(b), decreases from 3000 K at the substellar point to 50 K at $\pm 110^{\circ}$. Silicates start to melt at around $\pm 84^{\circ}$ with a temperature of 1700 K (solidus) and become fully molten at about $\pm 77^{\circ}$ with a temperature of 2000 K (liquidus) (Figure \ref{fig:thermal}(a)). The region within $-77^{\circ} \sim 77^{\circ}$ represents the surface magma ocean. Below surface, vertical diffusion gradually diffuses heat downward, increasing the interior temperature and melting the silicates (Figure \ref{fig:schematic}). 

\begin{figure*}
    \centering
    \includegraphics[width=0.95\textwidth]{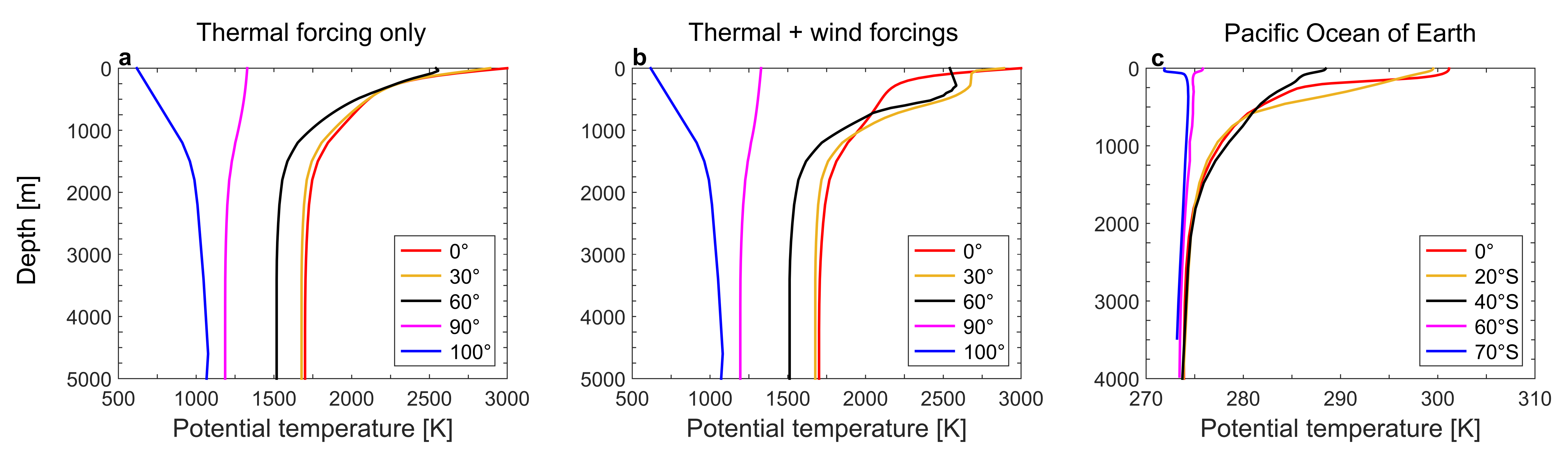}
    \caption{Comparison of vertical temperature profiles of this study with that in Earth's oceans. (a-b) Vertical temperature profiles at varying angles away from the substellar point under thermal forcing only (a) and under both thermal and wind forcings (b). (c) Vertical potential temperature profiles in the North Pacific of Earth at varying latitudes at 170$^{\circ}$ W.} 
    \label{fig:temp-profile}
\end{figure*}

Near the ocean edges, the temperature is nearly vertically uniform within the magma ocean. This arises from the narrow and intense vertical convection in that region (contours in Figure \ref{fig:thermal}(c)), induced by the relatively low temperature and high density. Silicates sink from the surface boundary, filling the bottom layer and equalizing the temperature at the bottom layer to that at the surface edges of the ocean (Figure \ref{fig:thermal}(a)). This phenomenon is similar to Earth's oceans\footnote{The long-term mean potential temperature data of Earth are from the NCEP Global Ocean Data Assimilation System (GODAS; \url{https://psl.noaa.gov/data/gridded/data.godas.html})}, in which the potential temperature in the deep ocean is determined by the surface temperature at high latitudes (Figure \ref{fig:temp-profile}(c); \cite{vallis2019essentials}). 
To compensate the strong downwelling motions near the edges, a wide range of weak upwelling motions occur around the substellar region. To pump the dense fluid back up to the surface consumes energy, which is provided by vertical diffusion.
The magma ocean reaches a steady state when the downward diffusion and upwelling motion balance with each other (right panel of Figure \ref{fig:schematic}). The stabilized magma ocean is bowl-shaped, with a maximum depth of $\sim$760 m at the substellar point (the black line in Figure \ref{fig:thermal}(a)).  

Beneath the magma ocean basin ($T< T_{liq}$), the temperature remains nearly vertically constant due to the dominant role of vertical diffusion. 
The small deviation from a vertically uniform temperature profile arises from the weak but non-zero circulation there (zonal velocity $u$ ranges from 10$^{-7}$ to 10$^{-4}$ m\,s$^{-1}$, figures not shown). Consequently, the thermal structure is governed by a balance between horizontal advection (horizontal diffusion) and vertical diffusion in regions with relatively large (small) velocity (for more details, refer to Appendix \ref{ap:timescale_momentum}). In particular, in solid regions with relatively large velocities, the interior temperature is underestimated due to horizontal advection (Figure \ref{fig:temp-profile}(a)). In contrast, in solid regions with relatively small velocities, the interior temperature is overestimated due to horizontal diffusion (the blue line of Figure \ref{fig:temp-profile}(a)).

It is noteworthy that the non-constant vertical temperature profile outside the magma ocean results from the numerical limitations of our 2D model. Firstly, we adopt an eddy diffusivity both horizontally and vertically, i.e., the horizontal diffusivity is significantly larger than the vertical diffusivity (Table \ref{tab1}). Secondly, the viscosity utilized below the solidus (10$^8$ m$^2$\,s$^{-1}$) is still significantly smaller than the realistic viscosity of solid silicates ($\sim$10$^{18}$ m$^2$\,s$^{-1}$; Section \ref{vis_dif}). Note that the interior temperature beyond the magma ocean might have not reached equilibrium due to the long timescales involved (Appendix \ref{ap:timescale_momentum}).
Despite this, the temperature beyond the magma ocean remains lower than the liquidus, which will not affect the above magma ocean. 


Silicate density is determined by temperature and pressure (Equation (\ref{equat10}) and Figure \ref{fig:model}(a)), and the influence of pressure is limited due to the shallow depths. Thus, silicate density follows a similar pattern to temperature, i.e., density is larger in the regions where temperature is low and is smaller in the regions where temperature is high (Figure \ref{fig:thermal}(b)). Note that there is a 10\% change in density from the solidus to the liquidus, which results in the largest density gradient near the phase transition regions.

In our 2D model without planetary rotation, zonal currents are primarily driven by zonal pressure gradient force and balanced by vertical viscosity (left panels of Figure \ref{fig:momentum}). Near the surface, the zonal pressure gradient force, dominated by sea surface height (SSH) gradient (Equation (\ref{equat7})), is positive (negative) on the east (west) of the substellar point (Figure \ref{fig:thermal}(d)). Thus, zonal currents near the surface flow outward from the substellar point, reaching a maximum value of $\sim$0.1 m\,s$^{-1}$ (colors in Figure \ref{fig:thermal}(c)). 
In the interior ocean, the zonal pressure gradient force, dominated by interior density gradient (Equation (\ref{equat7})), is negative (positive) on the east (west) of substellar point. Thus, interior zonal currents flow toward the substellar point. A clockwise (anti-clockwise) overturning circulation occurs on the east (west) of the substellar point, with upwelling motions around the substellar region and downwelling motions near the ocean edge (contours in Figure \ref{fig:thermal}(c)). Outside the ocean, ocean currents become weak due to the strong viscosity below the liquidus. 

SSH is positive within the magma ocean, reaching a maximum value of 30 m at the substellar point and gradually decreasing away from it, attaining a minimum value of $-$140 m at $\pm 110^{\circ}$ (Figure \ref{fig:thermal}(d)). 
As mentioned above, surface heat is gradually diffused downward over time, leading to an increase in the vertical-mean temperature and then a continuous rise in SSH within the ocean (Figure \ref{fig:time-series}(c)), due to thermal expansion. This corresponds to a persistent convergence of the vertical integral of zonal velocity within the ocean before the system reaches equilibrium (Equation (\ref{equat4})).
Conversely, outside the magma ocean, SSH experiences continuous decrease, becoming increasingly negative  (Figure \ref{fig:time-series}(c)). This corresponds to a continuous divergence of the vertical integral of zonal velocity before reaching equilibrium.

\begin{figure*}
    \centering
    \includegraphics[width=0.95\textwidth]{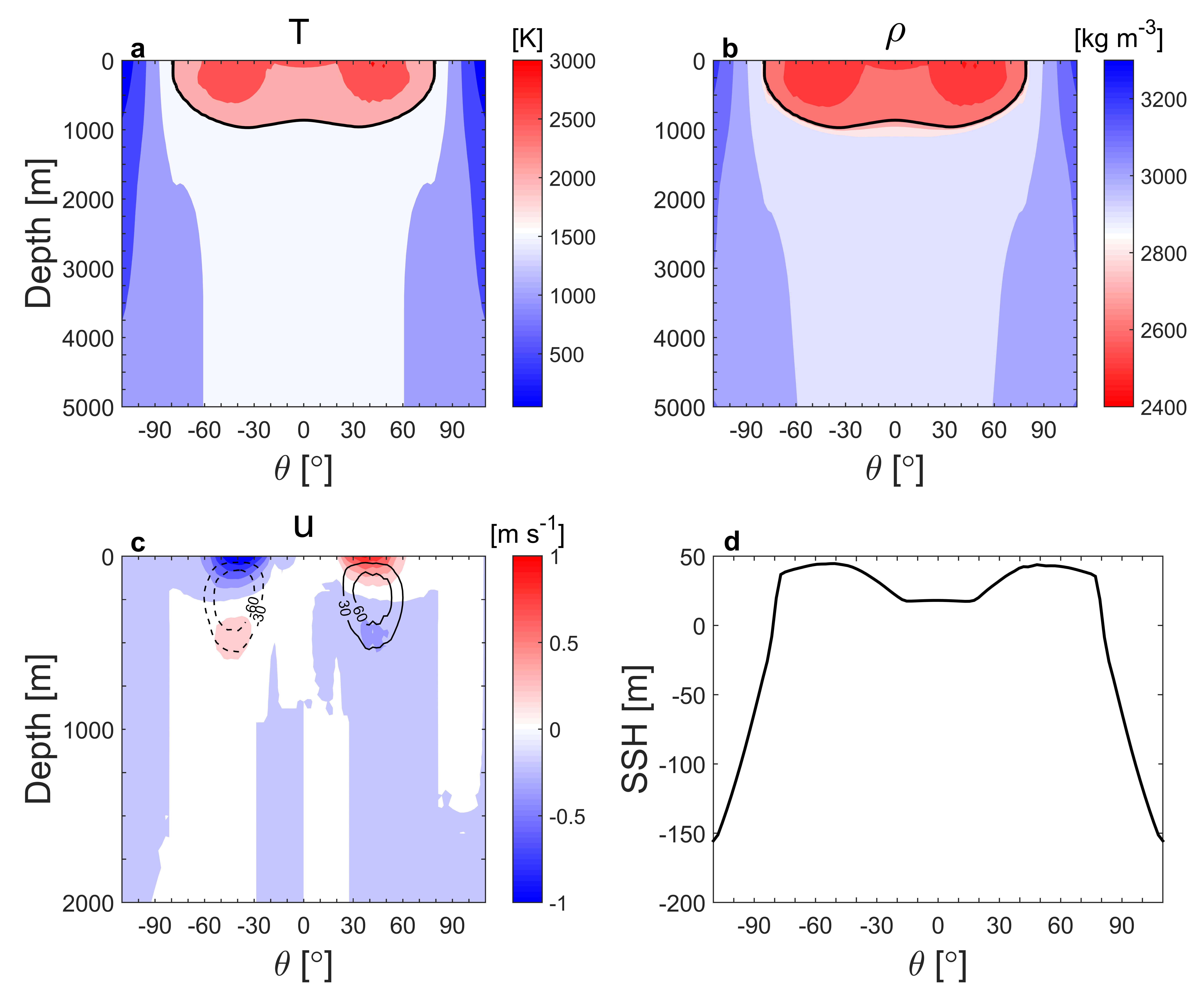}
    \caption{Same as Figure \ref{fig:thermal}, except that both thermal and wind forcings are included and A$_z$ = 10$^2$ m$^2$\,s$^{-1}$ at the liquidus in this simulation. The interval of velocity stream function is 30.0 m$^2$\,s$^{-1}$ in the contours of panel (c). }
    \label{fig:thermal_wind}
\end{figure*}

\subsection{Ocean Circulation under both Thermal and Wind Forcings}\label{wind}

Magma ocean depth remains several hundreds of meters deep, even with both thermal and wind forcings (Figure \ref{fig:thermal_wind}). The system reaches a dynamic equilibrium after $\sim$600 years of integration (Figure \ref{fig:time-series}(d-f)). The temperature field closely resembles that under thermal forcing alone (Figure \ref{fig:thermal_wind}(a)). In contrast, the ocean overturning circulation is primarily dominated by wind forcing. There are strong upwelling motions around $\pm 30^{\circ}$ and downwelling motions around $\pm 55^{\circ}$ induced by the divergence and convergence of zonal currents near the surface, respectively (Figure \ref{fig:thermal_wind}(c)). With the effect of wind stress, magma ocean becomes relatively shallower around the substellar region and relatively deeper around $\pm 40^{\circ}$ (Figure \ref{fig:schematic}). The magma ocean reaches its deepest depth of 960 m at $\pm 35^{\circ}$ (the black line in Figure \ref{fig:thermal_wind}(a)). 
The vertical temperature profiles under both thermal and wind forcings remain similar to that under thermal forcing only (Figure \ref{fig:temp-profile}(b)). 
The density pattern closely follows the temperature field and is primarily determined by temperature (Figure \ref{fig:thermal_wind}(b)).

In the presence of strong wind forcing, surface zonal currents are primarily driven by the zonal wind stress, which is balanced by vertical viscosity and the zonal pressure gradient force (right panels of Figure \ref{fig:momentum}). 
As a result, zonal currents align with the direction of zonal wind stress near the surface, flowing outward from the substellar point (Figure \ref{fig:model}(d) \& colors in Figure \ref{fig:thermal_wind}(c)). The currents reach a maximum velocity of 1.0 m\,s$^{-1}$ at approximately $\pm 40^{\circ}$, where the zonal wind stress is strongest. Below the surface layer, the contributions of SSH gradient and density gradient on the zonal pressure gradient are in the same direction, driving zonal currents toward the substellar point. To satisfy mass continuity, there is upwelling motion near $\pm 30^{\circ}$ and downwelling motion near $\pm 55^{\circ}$. Thus, a clockwise (anti-clockwise) overturning circulation forms on the east (west) side of the substellar point (contours in Figure \ref{fig:thermal_wind}(c)). 

SSH is immediately influenced by the imposed wind stress when the simulation starts. Thus, at the beginning of the simulation, SSH is extremely low around the substellar region due to the divergence of wind stress, and vice versa (Figure \ref{fig:time-series}(f)). As the magma ocean reaches deeper over time, SSH continuously increases due to thermal expansion, and vice versa outside the ocean. In the final equilibrated state, SSH exhibits a local minimum value of 15 m at the substellar point and a local maximum value of 45 m at around $\pm 60^{\circ}$ (Figure \ref{fig:thermal_wind}(d)).


\subsection{Comparisons with Previous Studies}\label{compare}
Magma ocean depth determined by ocean circulation is much shallower than that determined by adiabats (Figure \ref{fig:compare}). As in our study, when the interior heat source is weak or absent, the interior temperature is determined by stellar heating and ocean circulation. 
Along the magma ocean edge, fluid cools and sinks to the bottom, releasing gravity potential energy. To pump this dense fluid back up to the surface consumes energy, which is provided by vertical diffusion. In the upwelling regions near the substellar point, the downward diffusive heat flux needs to balance the upward advective heat flux, leading to the so called advective-diffusive balance (Appendix \ref{ap:timescale_momentum}; \cite{kite2016atmosphere,vallis2019essentials}).
\begin{equation}
w \frac{\partial T} {\partial z} \approx \frac{\partial} {\partial z} (k_z \frac{\partial T} {\partial z}).
\label{equat15}
\end{equation}
Thus, vertical temperature profile within the magma ocean can be given as,
\begin{equation}
T(z) = T_{b} + (T_s - T_{b}) e^{z/{\mathcal{D}}}, \mathcal{D} = k_z / w,
\label{equat16}
\end{equation}
where $T_s$ and $T_b$ represent the surface temperature and bottom temperature of the magma ocean, respectively, and $\mathcal{D}$ represents the characteristic scale for the thermocline depth, determined by the ratio of vertical diffusivity ($k_z$) to the upwelling velocity ($w$). In this study, the magma ocean boundary is determined by the liquids, i.e., $T_b = T_{liq}$. Furthermore, we assume that the magma ocean depth is comparable to the thermocline depth \citep{kite2016atmosphere}.
In the case of thermal forcing only (Figure \ref{fig:thermal}), with $k_z \sim$ 10$^{-4}$ m$^2$\,s$^{-1}$ and $w \sim 3\times10^{-7}$ m\,s$^{-1}$ within the magma ocean, the calculated thermocline depth is $\sim$300 m, being comparable to that obtained from the simulations. 


In scenarios when wind forcing dominates thermal forcing, the temperature profile deviates from advection-diffusion balance and is instead governed by horizontal and vertical advection (Appendix \ref{ap:timescale_momentum}; \cite{vallis2019essentials}),
\begin{equation}
u \frac{\partial T} {\partial x} + w \frac{\partial T} {\partial z} \approx 0, \mathcal{D}' = \frac{wL}{u},
\label{eq:temp_balance_wnd}
\end{equation}
where $\mathcal{D}'$ represents the characteristic scale for the thermocline depth in wind forcing-dominated scenario, and $L$ represents the horizontal scale for the magma ocean. In the simulation with both thermal and wind forcings (Figure \ref{fig:thermal_wind}), where $L \sim 10^7$ m, $u \sim 1.0$ m\,s$^{-1}$, and $w \sim 5\times10^{-5}$ m\,s$^{-1}$ within the magma ocean, the calculated thermocline depth is $\sim$500 m, aligning reasonably well with the results obtained from simulations.

Strictly speaking, the velocities utilized in Equations (\ref{equat16}) and (\ref{eq:temp_balance_wnd}) are intricately influenced by external forcings and various parameters, encompassing viscosity, diffusivity, gravity, thermal expansion coefficients, among others. However, for the sake of simplicity, we offer an estimation of magma ocean depth employing the zonal and vertical velocities derived from numerical simulations.
Referring to the theories in Earth's oceans \citep{vallis2019essentials}, \cite{kite2016atmosphere} provided scaling laws for horizontal velocity and magma ocean depth in scenarios dominated by thermal forcing. Nonetheless, their scaling analyses overlook the substantial impact of wind forcing. In Part II of this series of paper, we provide the dependences of both current speed and ocean depth on external forcings and diverse parameters across various dynamical regimes and forcing scenarios.

Magma ocean depth derived from our 2D model demonstrates a notable alignment with estimations derived from a geostrophic balance system (i.e., the pressure gradient force is balanced by the Coriolis force) by \cite{kite2016atmosphere}.  
Their scaling analysis indicated that the magma ocean depth in a geostrophic balance system is over 100 times shallower compared to scenarios without ocean circulation. 
In our 2D model, however, the zonal momentum is in balance between vertical viscosity term and the pressure gradient force (Figure \ref{fig:momentum}).
The consistency in magma ocean depth between models operating within varying dynamical regimes reinforces the pivotal influence of ocean circulation in modulating magma ocean depths. For a more comprehensive elucidation and comparative analysis across different dynamical regimes, please refer to Part II of this series of paper.


\subsection{Implications for observations}\label{observe}

\begin{figure*}
    \centering
    \includegraphics[width=0.99\textwidth]{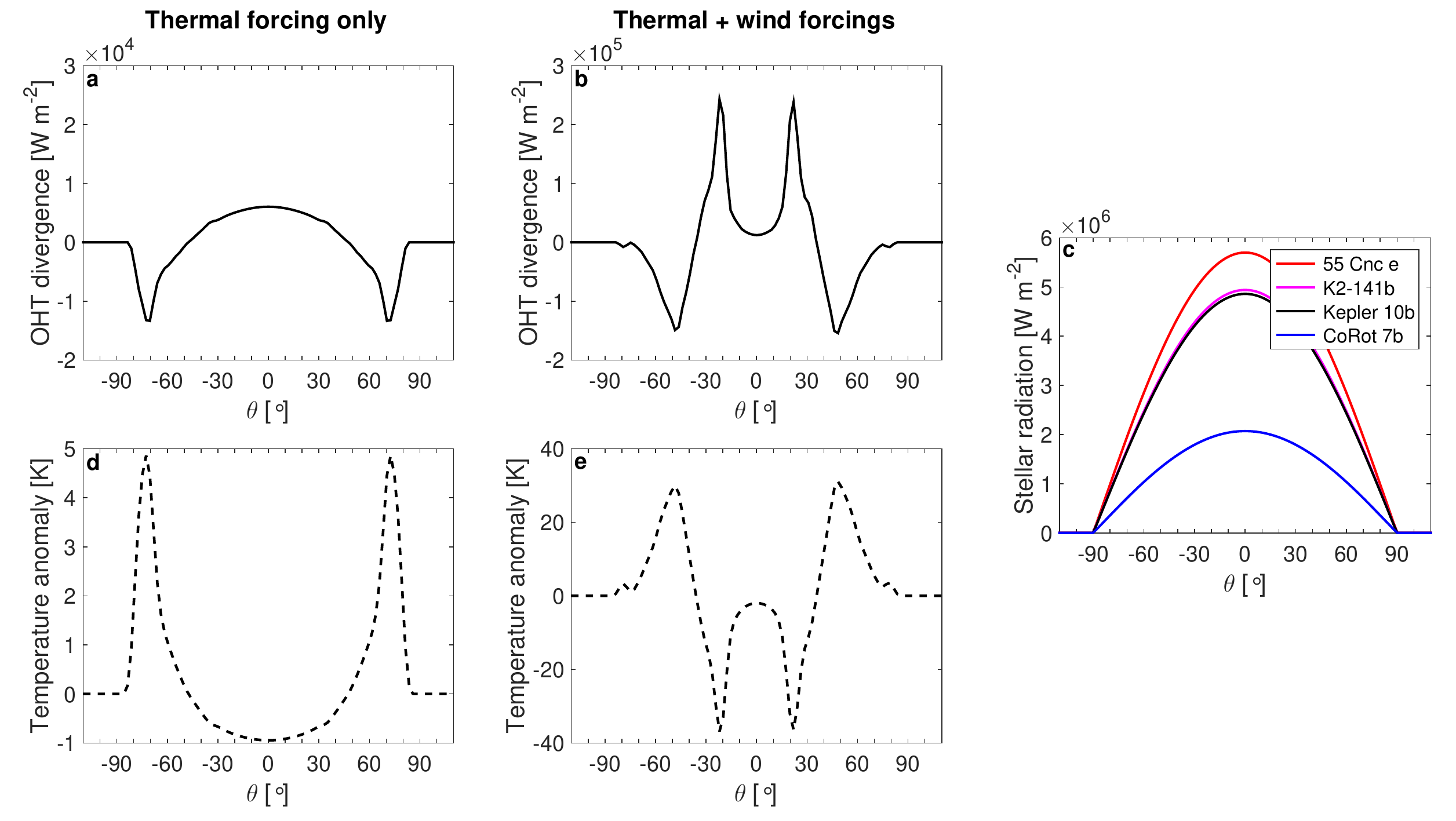}
    \caption{The ocean heat transport (OHT) divergence ($F_o = \frac{\partial} {\partial x} \int_{-\mathcal{D}}^0 {c_p \rho u T dz}$) (a \& b) and temperature anomaly induced by ocean circulation (d \& e) under thermal forcing only (left panels) and under both thermal and wind forcings (middle panels). Note that positive values of OHT divergence correspond to heat loss, and negative values of OHT divergence correspond to heat gain. Results shown in left and middle panels are based on the simulation results shown in Figures \ref{fig:thermal} and \ref{fig:thermal_wind}, respectively. The stellar insolation of different lava planets is shown for comparison (c). }
    \label{fig:oht}
\end{figure*}

\cite{kite2016atmosphere} estimated the ocean heat transport (OHT) divergence due to thermal-driven overturning circulation on lava planets. Using simple scalings based on 3D equations, they suggested an oceanic current speed of 10$^{-2}$ m\,s$^{-1}$ and a magma ocean depth of $\mathcal{O}$(100)  m (consistent with Figure \ref{fig:thermal}). Consequently, their calculated OHT divergence was approximately four orders of magnitude lower than the stellar insolation  ($\sim$10$^6$ W\,m$^{-2}$; Figure \ref{fig:oht}(c)). \cite{leger2011extreme} also provided an estimation for OHT divergence on CoRoT-7b. In contrast to the geostrophic balance system, they considered zonal current speed without the Coriolis force and attained a maximum value of $\sim$2 m\,s$^{-1}$. With an ocean depth of 100 m, they suggested an OHT convergence of 10$^4$ W\,m$^{-2}$, still two orders of magnitude smaller than the stellar insolation. 
 
Here, the OHT divergence of numerical simulations with thermal forcing alone and with both thermal and wind forcings are shown in Figure \ref{fig:oht}. 
If the velocity were entirely zero in solid regions, there would be no ocean heat transport. Thus, only the OHT divergence within the magma ocean is calculated.
For comparison, the stellar insolations of four different tidally locked lava planets are presented in Figure \ref{fig:oht}(c). In the case of thermal forcing alone (Figure \ref{fig:oht}(a)), OHT diverges within about $\pm 50^{\circ}$ and converges outside $\pm 50^{\circ}$. 
The OHT divergence/convergence reaches a maximum value of $\sim$10$^4$ W\,m$^{-2}$ under a substellar temperature of 3000 K, nearly two orders of magnitude lower than the stellar insolation (Figure \ref{fig:oht}(c)). 
With both thermal and wind forcings (Figure \ref{fig:oht}(b)), the OHT divergence/convergence reaches a maximum value of $2 \times 10^5$ W\,m$^{-2}$, remaining smaller than the stellar insolation. 

To quantify the effects of ocean circulation on surface temperature, we estimate the ocean temperature anomaly ($T'$) induced by ocean circulation using the following formula,
\begin{equation}
T' = \left\{
\begin{array}{rcl}
(\frac{S-F_o}{\sigma_b})^{\frac{1}{4}} - (\frac{S}{\sigma_b})^{\frac{1}{4}}, & &  {S \geq F_o}\\
-(\frac{F_o -S}{\sigma_b})^{\frac{1}{4}} - (\frac{S}{\sigma_b})^{\frac{1}{4}}, & &  {S < F_o}
\end{array}\right.
\label{eq:oht_anomaly}
\end{equation}
where $S$ represents stellar insolation, $F_o$ represents the OHT divergence, and $\sigma_b$ represents the Stefan–Boltzmann constant. It should be noted that positive values of $F_o$ correspond to heat loss, leading to a negative temperature anomaly, while negative values of $F_o$ correspond to heat gain, resulting in a positive temperature anomaly. 
In the case of thermal forcing alone, the temperature anomaly ($T'$) induced by ocean circulation ranges from $-1$ to 5 K (Figure \ref{fig:oht}(d)). In particular, the ocean circulation decreases the temperature within  $\pm50^{\circ}$, while increases the temperature around $\pm70^{\circ}$.
When both thermal and wind forcings are imposed, $T'$ ranges from $-40$ to  20 K (Figure \ref{fig:oht}(e)). Specifically, the ocean circulation decreases (increases) the temperature around $\pm20^{\circ}$ ($\pm50^{\circ}$). 
Overall speaking, the ocean circulation tends to reduce the horizontal temperature gradient within the magma ocean. However, the temperature anomaly remains significantly lower than the radiative equilibrium temperature. Consequently, the impact of oceanic overturning circulation on surface temperature, and thus on the thermal phase curve, is limited.

The OHT divergence increases with the intensity of stellar insolation. As the stellar radiation intensifies, both the substellar temperature and the horizontal scale of the magma ocean increase, leading to stronger ocean circulation and deeper magma ocean. Consequently, the OHT divergence escalates. However, even under varying intensities of stellar radiation, the OHT divergence remains two orders of magnitude smaller than the stellar insolation. For further insights, we delve into detailed numerical and theoretical analyses in Part II of this series of paper.

\section{Discussions}\label{sec:discuss}

\subsection{Influence of latent heat}\label{sec:discuss_latent-heat}
The ocean model has not been coupled with the atmosphere. Surface temperature is completely determined by local radiative balance \citep{leger2009transiting}. As mentioned in Section \ref{sec:intro}, atmospheric latent heat is much smaller than the stellar insolation \citep{castan2011atmospheres,kite2016atmosphere}. 
This assumption seems to work for an evaporation-driven atmosphere \citep{ingersoll1985supersonic}. 


\subsection{Influence of molecular diffusivity}\label{sec:discuss_kappa}
The variation of silicate diffusivity with temperature is not included in our 2D model. Basically, silicate molecular diffusivity can vary approximately from 10$^{-20}$ to 10$^{-8}$ m$^2$\,s$^{-1}$ from solid to liquid states (Section \ref{vis_dif}). In our 2D model, a constant and globally uniform eddy diffusivity is employed for simplicity, which is significantly larger than the molecular diffusivity. Outside the magma ocean, the temperature is governed by vertical diffusion. Thus, the magnitude of diffusivity will not exert an evident influence on the thermal structure, which will primarily affect the equilibrium time (Appendix \ref{ap:timescale_momentum}). However, the significant difference between the horizontal diffusivity and vertical diffusivity allows the horizontal diffusion to play a role in affecting the thermal structure in solid regions (Sections \ref{thermal} \& \ref{wind}).
%


\subsection{Influence of planetary rotation}\label{sec:discuss_rotate}
The 2D model does not account for the Coriolis force. The significance of planetary rotation can be evaluated using the Rossby number ($R_o = \frac{u}{\Omega L}$, where $L$ is the typical scale of the horizontal range of the magma ocean and $\Omega$ is the planetary rotation rate) and the Ekman number ($E_z = \frac{A_z}{\Omega D^2}$, where $D$ is the typical scale of the vertical range of the magma ocean) \citep{vallis2019essentials}. Here, $R_o$ represents the ratio of the nonlinear advection term to the Coriolis force, while $E_z$ denotes the ratio of the vertical viscosity term to the Coriolis force.

Using the planetary rotation rate of Kepler-10b, $\Omega \approx \frac{2\pi}{(0.84\,days)} \sim 10^{-4}$ s$^{-1}$ \citep{batalha2011kepler}, and assuming that $L \sim 10^7$ m, $D \sim 10^3$ m, $A_z \sim 10^1$ m$^2$\,s$^{-1}$, and $u \sim 0.1$ m\,s$^{-1}$ for thermal forcing-dominated simulation (Figure \ref{fig:thermal}), we can calculate Rossby number $R_o \sim 5\times 10^{-4}$ and Ekman number $E_z \sim 0.1$. Hence, both the nonlinear advection and vertical viscosity terms are significantly smaller than the Coriolis force, spanning 1--4 orders of magnitude. Consequently, if the Coriolis force were considered, the system would likely be in geostrophic balance, in which the pressure gradient force is balanced by the Coriolis force.


In Part II of this series of paper, we delve into a detailed discussion and comparison of the ocean circulation and ocean depth across three different dynamical regimes: viscosity-dominant, advection-dominant, and rotation-dominant regimes. Theoretical results suggest that the difference in magma ocean depth among the three dynamical regimes is relatively constrained, consistently shallower than that determined by the adiabat.

\subsection{Influence of internal heating}\label{sec:discuss_interna-heat}
Given that most tidally locked lava planets have cooled over billions of years, internal heating is not fully considered in our 2D model. 
In the absence of ocean circulation, the interior temperature is governed by diffusion. Without internal heating, the interior temperature should be vertically uniform. By comparison, the interior temperature gradually increases from the surface temperature with depth in the presence of internal heating. 
The vertical temperature gradient ($\gamma$) induced by internal heating is expressed as \citep{scott2001,adcroft2001,Emile2009,Wang2016},
\begin{equation}
\gamma = -\frac{\partial T}{\partial z} = \frac{Q}{\rho c_p k_z},
\label{eq:internal_heating}
\end{equation}
where $Q$ represents the internal heating rate and $z$ represents the distance from the surface. 
Assuming an internal heating rate of $Q \sim 0.5$ W\,m$^{-2}$ \citep{leger2011extreme}, along with the ocean parameters presented in Table \ref{tab1}, the resulting vertical temperature gradient $\gamma \sim 1$ K\,km$^{-1}$. This implies that the temperature would increase from 3000 K at the surface to 3100 K at a depth of 100 km near the substellar point. Thus, when internal heating is weak, its effect on the magma ocean is not strong. However, when internal heating is so strong that vigorous convection occurs, magma ocean depth could be significantly increased (Figure \ref{fig:compare}(a); \cite{leger2011extreme,boukare2022deep,boukare2023,meier2023}). 


\subsection{Surface boundary of magma ocean}\label{sec:discuss_boundary}

In this study, we assume a fixed magma ocean boundary at the sea surface, because surface temperature is fixed in the simulations. This assumption proves effective for tidally locked lava planets, where the influence of ocean circulation on both surface temperature and magma ocean boundary is negligible (Figure \ref{fig:oht}). Furthermore, we can also discern which process dominates the surface temperature by comparing the timescales of radiative equilibrium, horizontal diffusion, and horizontal advection (Equation (\ref{eq:timescale_T})). The radiative timescale can be estimated using the formula \citep{showman2009atmospheric}: $\tau_{rad} = \frac{c_p p_s}{4 g \sigma_b T_e^3}$, where $p_s$ denotes the atmospheric pressure, and $T_e$ denotes the radiative equilibrium temperature. For the silicate atmosphere on tidally locked lava planets \citep{castan2011atmospheres,nguyen2020modelling,kang2021escaping}, assuming $c_p \sim 1800$ J\,kg$^{-1}$\,K$^{-1}$, $p_s \sim 10^4$ Pa, and $g \sim 20$ m\,s$^{-2}$, the radiative equilibrium timescale is $\sim$400 s near magma ocean boundary ($T_e \sim 2000$ K).  Given that zonal velocity is approximately $10^{-4}$ m\,s$^{-1}$ there (figure not shown),  both the timescales of horizontal diffusion and horizontal advection are around 3000 years (Appendix \ref{ap:timescale_momentum}), which are considerably longer than the radiative timescale. Overall, horizontal diffusion and advection are unable to extend the magma ocean boundary beyond the day/night terminator.  


\subsection{Mass-driven overturning circulation}\label{sec:discuss_mass-driven}
For a tidally locked lava planet, there is persistent evaporation (condensation) at the magma surface where the atmospheric pressure is lower (higher) than the saturated vapor pressure \citep{castan2011atmospheres,nguyen2020modelling}. Therefore, there is mass loss where surface evaporates and mass gain where atmosphere condenses. For example, for a Na atmosphere on K2-141b, magma surface loses mass within about $\pm 40^{\circ}$ and gains mass elsewhere \citep{nguyen2020modelling}. The horizontal gradient in mass flux can result in gradient in surface height and thereby in surface pressure, which then drives an overturning circulation. 
Different from the thermal-driven or wind-dominated circulation (Figures \ref{fig:thermal} \& \ref{fig:thermal_wind}), the mass-driven circulation exhibits an opposite direction, with surface ocean currents flowing toward the substellar point \citep{tokano2016sun,kite2016atmosphere}. This mass-driven circulation is beyond the scope of our study owing to the limitations of our 2D model. And its influences on the surface energy budget and surface composition need to be addressed using a more complicated 3D ocean general circulation model.


\section{Summary}\label{sec:conclude}

Magma ocean can form and persist in the dayside of tidally locked lava planets, where surface temperature are high enough to melt typical silicates under intense stellar radiation. In this study, we aim to simulate the magma ocean of tidally locked lava planets using an idealized 2D (x-z) ocean circulation model under the assumption of a small or no internal heat flux. Our main results are as follows:

(1) In the presence of thermal-driven overturning circulation, magma ocean depth is only $\mathcal{O}$(100) m. The overturning circulation, driven by surface temperature contrast, is clockwise (anti-clockwise) on the east (west) of the substellar point.

(2) The zonal wind stress can reach $\sim$100 N\,m$^{-2}$, roughly 1000 times that on Earth. Under such strong wind forcing, the ocean overturning circulation intensifies. Overall, magma ocean depth under thermal and wind forcings remains $\mathcal{O}$(100)  m.

(3) Determined by ocean circulation, magma ocean depth is more than 100 times shallower than that determined by adiabats. Our simulation results are consistent with that suggested by \cite{kite2016atmosphere}.

(4) The ocean heat transport divergence/convergence is 1--2 orders of magnitude smaller than the stellar radiation. 
Thus, the effects of ocean circulation on the climate are limited. 

In Part I of this program, we presented simulation results for ocean circulation and ocean depth in a non-rotating, viscosity-dominant 2D system.
In Part II, we will theoretically explore ocean circulation and ocean depth in three different dynamical regimes: non-rotating viscosity-dominant, non-rotating inviscid limit, and rotation-dominant regimes (Lai, Kang, \& Yang: Ocean Circulation on Tide-locked Lava Worlds, Part II: Scalings). We further compare the scaling results with those from numerical simulations. The results consistently indicate that, under different mechanisms and parameters, magma ocean depth in the presence of ocean circulation remains $\mathcal{O}$(100)  m.   

\section*{Acknowledgments}
We express our gratitude to Feng Ding, Xinyu Wen, and Yonggang Liu for their valuable contributions and insightful discussions. J.Y. is supported by NSFC under grant nos. 42075046, 42275134 and 42161144011. This work has been supported by the science research grants from the China Manned Space Project (No. CMS-CSST-2021-B09). W.K. is supported by the MIT startup fund. 

The 2D model employed in this study is developed by the authors and is available at \url{https://github.com/YanhongLai/Two-dimensional-model-for-magma-ocean-on-lava-worlds.git}. The simulation data used are archived at \url{https://doi.org/10.5281/zenodo.11467142}.

\begin{appendix}
\setcounter{equation}{0}
\renewcommand\theequation{\Alph{section}\arabic{equation}}
\renewcommand\thefigure{\Alph{section}\arabic{figure}}
\setcounter{figure}{0}

\section{Time series and momentum budget}\label{ap:timescale_momentum}

\begin{figure*}
    \centering
    \includegraphics[width=0.9\textwidth]{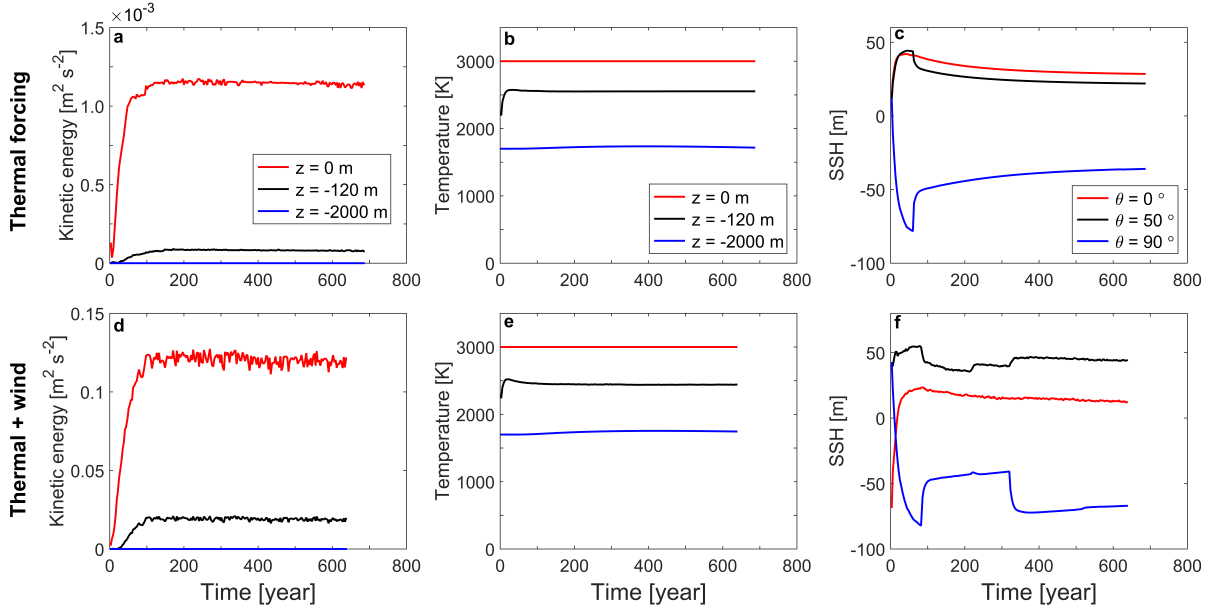}
    \caption{Time series of the simulations shown in Figure \ref{fig:thermal} (upper panels) and Figure \ref{fig:thermal_wind} (lower panels). Zonal-mean kinetic energy at different depths (a \& d), temperature of the substellar point at varying depths (b \& e), and sea surface height (SSH) at different angles (c \& f), as a function of time. Note that the system is initialized with a rest state.}
    \label{fig:time-series}
\end{figure*}

Numerical simulations approximately reach dynamical equilibrium after 600 years of integration. For example, the time series of zonal-mean kinetic energy, temperature at the substellar point, and SSH are presented in Figure \ref{fig:time-series}. Initialized from a rest state, zonal-mean kinetic energy increases over time. It reaches maximum values of about 10$^{-3}$ m$^2$\,s$^{-2}$ at the surface for simulations with thermal forcing only and 0.12 m$^2$\,s$^{-2}$ when both thermal and wind forcings are included (Figure \ref{fig:time-series}(a \& d)). 
Note that oscillations in the kinetic energy time series may be induced by numerical instability. 

The time required for equilibrium can be estimated by simple scalings. According to Equations (\ref{equat1}) and (\ref{equat2}), the equilibrium time for zonal velocity (kinetic energy) and temperature can be given as, respectively,
\begin{equation}\label{eq:timescale_u}
\tau_u \sim max\{ \frac{L}{U}, \frac{H}{W}, \frac{L^2}{A_h}, \frac{H^2}{A_z}\},
\end{equation}
\begin{equation}\label{eq:timescale_T}
\tau_T \sim max\{ \frac{L}{U}, \frac{H}{W}, \frac{L^2}{k_h}, \frac{H^2}{k_z}\},
\end{equation}
where $L$ and $H$ are the horizontal and vertical scales of the simulated domain, respectively; $U$ and $W$ are the characteristic values of zonal and vertical current speeds, respectively. Given the significant difference in the viscosity and current speed within and outside the magma ocean, we estimate their timescales of each term separately. 

The following estimation takes the results under both thermal and wind forcings as an example (Figure \ref{fig:thermal_wind}). Within the magma ocean, $U \sim$ 1.0 m\,s$^{-1}$, $W \sim$ 5$\times$10$^{-5}$ m\,s$^{-1}$, $A_h$ $\sim$ 10$^4$ m$^2$\,s$^{-1}$, $A_z$ $\sim$ 10$^2$ m$^2$\,s$^{-1}$, $k_h$ $\sim$ 10$^3$ m$^2$\,s$^{-1}$, $k_z$ $\sim$ 10$^{-4}$ m$^2$\,s$^{-1}$, $L \sim$ 2$\times$10$^{7}$ m, and $H \sim$ 5$\times$10$^3$ m, we can estimate that $\tau_u \sim 300$ years and $\tau_T \sim 3000$ years, which are restricted by horizontal viscosity and horizontal diffusion, respectively. In particular, the temperature is actually dominated by horizontal and vertical advection within the magma ocean (for details, see below). Thus, the temperature within the magma ocean can reach dynamic equilibrium after $\sim$300 years of integration.
Outside the magma ocean, we take the results in the regions where the temperature is around the solidus as an example: $U \sim$ 10$^{-5}$ m\,s$^{-1}$, $W \sim$ 10$^{-8}$ m\,s$^{-1}$, $A_h$ $\sim$ 10$^8$ m$^2$\,s$^{-1}$, and $A_z$ $\sim$ 10$^8$ m$^2$\,s$^{-1}$, the values of both $\tau_u$ and $\tau_T$ are $\sim$30,000 years, restricted by horizontal advection. 

\begin{figure*}
    \centering
    \includegraphics[width=0.99\textwidth]{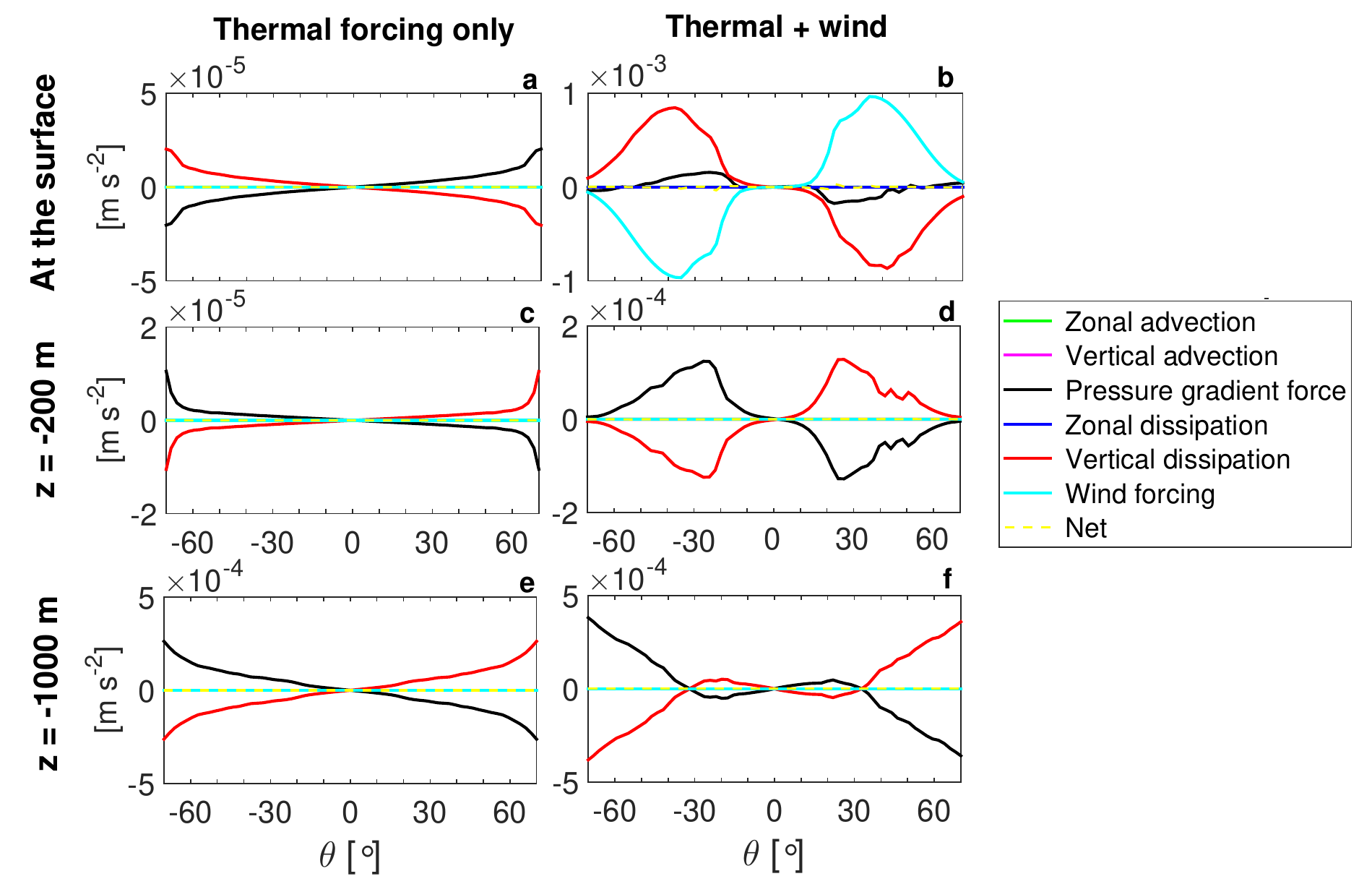}
    \caption{The zonal momentum budget at different depths of the simulations shown in Figure \ref{fig:thermal} (left panels) and Figure \ref{fig:thermal_wind} (right panels). From top to bottom panels, they are zonal momentum budget at the surface (a \& b), at a depth of 200 m (c \& d), and at a depth of 1000 m (e \& f), respectively. Due to the substantial difference in zonal momentum between the interior and exterior magma ocean, only the zonal momentum budget within $\pm70^{\circ}$ is shown.}
    \label{fig:momentum}
\end{figure*}

The timescales of each term in the momentum and thermal equations not only determine the time required to reach equilibrium, but also play a crucial role in determining the dominant physical processes governing the ocean dynamics. Different from the equilibrium time, the ocean dynamics is governed by terms with relatively short timescales. In the zonal momentum equation, the timescale of the vertical viscosity term is consistently smallest everywhere, which dominates over the advection and horizontal viscosity terms. Thus, the ocean circulation is governed by a balance between vertical viscosity and pressure gradient force (Figure \ref{fig:momentum}).  

In the thermal equation, the dominant physical processes differ within and outside the magma ocean. Within the magma ocean ($T \ge T_{liq}$), both the timescales of horizontal advection and vertical advection are approximately 0.3 years, which are shorter than remaining terms. Outside the magma ocean ($T< T_{liq}$), $U$ ranges from 10$^{-7}$ to 10$^{-4}$ m\,s$^{-1}$, and $W$ ranges from 10$^{-9}$ to 10$^{-7}$ m\,s$^{-1}$. Thus, in solid regions with relatively large velocities, the timescales of horizontal advection and vertical diffusion are smallest and comparable ($\sim$3000 years). By comparison, in solid regions with relatively small velocities, the timescales of horizontal diffusion and vertical diffusion are smallest and comparable ($\sim$3000 years). 

According to the above estimation, under both thermal and wind forcings, the thermal structure is governed by a balance between horizontal advection and vertical advection within the magma ocean. Outside the magma ocean, the thermal structure is governed by a balance between horizontal advection (horizontal diffusion) and vertical diffusion in regions with relatively large (small) velocity. It is noteworthy that under thermal forcing only (Figure \ref{fig:thermal}), the thermal structure within the magma ocean is controlled by the balance between vertical advection and vertical diffusion .

In the 2D model, due to the lack of the Coriolis force, the system is not in geostrophic balance as that in 3D models \citep{kite2016atmosphere,vallis2019essentials}. Without wind forcing, zonal momentum is in balance between the zonal pressure gradient force and vertical viscosity both at the surface and in the interior ocean (left panels of Figure \ref{fig:momentum}). For example, on the east of the substellar point: the zonal pressure gradient force is determined by SSH gradient and is positive at the surface, while it is determined by the interior density gradient and is negative in the interior. Thus, zonal currents in the upper and lower ocean are in opposite directions (Figure \ref{fig:thermal}(c)).  In the presence of strong wind forcing, the wind forcing is balanced by the zonal pressure gradient force and vertical viscosity at the surface (Figure \ref{fig:momentum}(b)). The zonal momentum remains in balance between the zonal pressure gradient force and vertical viscosity below the surface (Figure \ref{fig:momentum}(d \& f)). 

\end{appendix}
\bibliography{main}{}
\bibliographystyle{aasjournal}

\end{document}